\documentclass{aa}
\usepackage{graphicx}
\usepackage{amsmath}
\usepackage{txfonts}

\newcommand{\mach}{${\mathcal M}$}
\newcommand{\sbar}{$\bar{s}$}
\newcommand{\mbar}{$\bar{m}$}

\begin{document}

  \title{Evolving structures of star-forming clusters}

  \author{S. Schmeja
         \and
         R. S. Klessen 
	  }


  \institute{Astrophysikalisches Institut Potsdam, An der Sternwarte 16,
             14482 Potsdam, Germany\\
              \email{sschmeja@aip.de, rklessen@aip.de}
            }

  \date{Received ; accepted }

  \abstract
   {Understanding the formation and evolution of young star clusters requires quantitative
    statistical measures of their structure.}
   {We investigate the structures of observed and modelled star-forming clusters.
    By considering the different evolutionary classes in the observations and the temporal
    evolution in models of gravoturbulent fragmentation, we study the temporal evolution
    of the cluster structures.}
   {We apply different statistical methods, in particular the normalised mean correlation length
    and the minimum spanning tree technique.
    We refine the normalisation of the clustering parameters by defining the area using the
    normalised convex hull of the objects and investigate the effect of two-dimensional projection of
    three-dimensional clusters.
    We introduce a new measure $\xi$ for the elongation of a cluster. It is
    defined as the ratio of the cluster radius determined by an enclosing
    circle to the cluster radius derived from the normalised convex
    hull.
    }
   {The mean separation of young stars increases with the evolutionary class, reflecting
   the expansion of the cluster.
   The clustering parameters of the model clusters correspond in many cases well
   to those from observed ones, especially when the $\xi$ values are similar.
   No correlation of the clustering parameters with the turbulent environment of the molecular cloud
   is found, indicating that possible influences of the environment on the clustering
   behaviour are quickly smoothed out by the stellar velocity dispersion.
   The temporal evolution of the clustering parameters shows that the star cluster
   builds up from several subclusters and evolves to a more centrally concentrated cluster,
   while the cluster expands slower than new stars are formed.}
   {}


  \maketitle
%

\section{Introduction}

Almost all stars form in clusters. Embedded clusters contain various types of
young stars, making them ideally suited to study the early stages of star formation as they
provide a large and genetically homogeneous sample (see Lada \& Lada
\cite{lada_lada} for a review).
Understanding the formation and evolution of young stellar clusters requires
quantitative statistical measures of their structure, which may give
important clues to the formation process.
While some clusters are centrally concentrated with a smooth radial
density gradient, others show filaments and signs of fractal subclustering.
If and how different structures are connected to the environmental conditions
of the molecular clouds and how they depend on the evolutionary stage of
the cluster is not yet clear.

Different methods have been used to describe the clustering properties of star clusters, e.g.,
the mean surface density of companions (Larson \cite{larson95}) or spanning trees.
Cartwright \& Whitworth (2004; hereafter \cite{cw04}) presented a review of various statistical methods
for analysing the structures of star clusters and a detailed investigation of both observed and
artificially created clusters.
We apply the methods discussed there to clusters created by numerical simulations of gravoturbulent star formation
and investigate the clustering behaviour with time. We extend the analysis of observed clusters
by considering different evolutionary classes.

Four classes of young stellar objects (YSOs) are distinguished
according to the properties of their spectral energy distributions
(SEDs) (e.g. Andr\'e et al. \cite{andre00}):
Class\,0 sources are deeply embedded protostars with a large sub-mm to
bolometric luminosity ratio (L$_{\rm smm}$/L$_{\rm bol} >$\,0.005).
The Class\,0 stage is the main accretion phase and lasts only a few
$10^4$\,yr. Class\,1 objects are relatively evolved protostars, which
are surrounded by an accretion disc and a circumstellar envelope.
Pre-main-sequence stars in Class\,2 and 3 correspond to classical
and weak line T~Tauri stars, respectively. They are characterised by a
circumstellar disc (optically thick in Class\,2, optically thin in
Class\,3) and the lack of a dense circumstellar envelope.
These four classes are usually interpreted as an evolutionary sequence
from Class\,0 to 3.
The progenitors of these forming stars are prestellar cores (starless
cores, prestellar condensations). These are gravitationally bound,
dense molecular cloud cores with typical stellar masses that may
already be in a state of collapse, but have not formed a central
protostellar object yet.

Sect.~\ref{sec:statistics} explains the statistical methods used. In Sect.~\ref{sec:obsdata} and \ref{sec:models} we describe the observations and the models, respectively, and
in Sect.~\ref{sec:application} the application of the methods to the data.
The results are analysed and discussed in Sect.~\ref{sec:discussion}, while we present our conclusions in Sect.~\ref{sec:conclude}.
The Appendices give the details about the normalisation of the clustering parameters.

   \begin{table*}[t]
\caption{Clustering measures and numbers for the observed star-forming regions for prestellar cores (p), Class~1, 2, 3 sources, and all YSOs. (Note that the total number of YSOs is larger
than the sum of objects in the individual classes, because it contains Class~0 sources and objects
with unclear classification not considered in the analysis of the individual classes.)
For details see the discussion in Sect.~\ref{sec:discuss_obs}.
}
 \label{tab:obs_results}

\begin{center}
\begin{tabular}{l r r r r r r r r r r r r r }
\hline
\hline
Region & Class & $n$ & s [$'$] & s [pc] & m [$'$] & m [pc] & $\bar{s}$
             & $\bar{m}$ & $Q$  & $\bar{s}^*$ & $\bar{m}^*$ & $Q^*$ & $\xi$\\
\hline
$\rho$ Ophiuchi & p & 98 & 20.45 & 0.83 & 2.71 & 0.11 & 0.67 & 0.55 & 0.82 & 0.49 & 0.40 & 0.81 & 1.36 \\
          & 1 & 15 & 11.82 & 0.48 & 3.15 & 0.13 & 1.13 & 0.68 & 0.60 & 0.77 & 0.43 & 0.56 & 1.47 \\
          & 2 & 111& 15.86 & 0.65 & 2.25 & 0.09 & 0.69 & 0.58 & 0.83 & 0.54 & 0.63 & 1.15 & 1.27 \\
          & 3 & 77 & 16.25 & 0.66 & 2.53 & 0.10 & 0.82 & 0.63 & 0.76 & 0.70 & 0.66 & 0.93 & 1.17 \\
          & all & 205 & 15.63 & 0.64 & 1.67 & 0.07 & 0.67 & 0.57 & 0.85 & 0.53 & 0.45 & 0.85 & 1.28 \\
\hline
Serpens   & 1 & 19 & 3.43 & 0.26 & 1.12 & 0.09 & 0.99 & 0.82 & 0.83 & 0.27 & 0.22 & 0.79 & 3.61 \\
          & 2 & 18 & 7.85 & 0.59 & 1.74 & 0.13 & 0.74 & 0.61 & 0.82 & 0.52 & 0.42 & 0.80 & 1.42 \\
          & all & 80 & 5.98 & 0.45 & 1.06 & 0.08 & 0.60 & 0.54 & 0.90 & 0.37 & 0.32 & 0.89 & 1.63 \\
\hline
Taurus    & p & 52 & 281.0 & 11.47 & 45.88 & 1.87 & 0.77 & 0.48 & 0.62 & 0.59 & 0.36 & 0.61 & 1.29 \\
          & 1 & 25 & 265.6 & 10.84 & 73.18 & 2.98 & 0.83 & 0.60 & 0.73 & 0.59 & 0.41 & 0.70 & 1.39 \\
          & 2 &108 & 333.5 & 13.62 & 34.30 & 1.40 & 0.69 & 0.41 & 0.59 & 0.43 & 0.54 & 1.26 & 1.61 \\
          & 3 & 72 & 414.4 & 16.96 & 52.99 & 2.16 & 0.84 & 0.51 & 0.60 & 0.77 & 0.63 & 0.82 & 1.09 \\
          & all & 197 & 356.8 & 14.58 & 25.98 & 1.06 & 0.67 & 0.39 & 0.58 & 0.46 & 0.27 & 0.58 & 1.45 \\
\hline
IC 348    & all & 288 & 6.86 & 0.63 & 0.66 & 0.06 & 0.56 & 0.52 & 0.93 & 0.49 & 0.45 & 0.92 & 1.16 \\
\hline
Chamaeleon~I & all & 180 & 41.51 & 1.81 & 3.74 & 0.16 & 0.55 & 0.37 & 0.68 & 0.50 & 0.34 & 0.68 & 1.32 \\
\hline
\end{tabular}
\end{center}
\end{table*}

\section{Statistical Methods}
\label{sec:statistics}

\subsection{Normalised correlation length}

A wide range of statistical methods has been developed to analyse the structure of star clusters
(see \cite{cw04} for a review).
A simple approach is to study the distribution of source separations, as it has been
done e.g. by Kaas et al.\ (\cite{kaas04}) for the Serpens cloud core.
Larson (\cite{larson95}), extending the analysis by Gomez et al.\ (\cite{gomez93}),
introduced the mean surface density of companions $\Sigma (\theta)$, a
tool that since then has often been used to study star-forming clusters (e.g. Bate et al.\
\cite{bcm98}; Gladwin et al.\ \cite{gladwin99}; Klessen \& Kroupa \cite {klessen+kroupa}).
The mean surface density of companions (MSDC) specifies the average number of neighbours per square degree
on the sky at an angular separation $\theta$ for each cluster star. Knowing the distance to the cluster,
$\theta$ can be converted to an absolute distance $r$ to determine $\Sigma (r)$.
The physical interpretation of the MSDC can be difficult, and \cite{cw04} have shown
that the normalised correlation length is a better indicator for the clustering behaviour.
The normalised correlation length $\bar{s}$ is the mean separation $s$ between stars in the cluster,
normalised by dividing by the radius of the cluster, $R_{\rm cluster}$.
This radius is defined via the normalised convex hull of the objects (see Appendix~\ref{app:area}).
The $\bar{s}$ values are independent of the number of stars in the cluster (\cite{cw04}).

\subsection{Minimum spanning trees}

The minimum spanning tree (MST), a construct from graph theory, is the unique set of straight
lines (``edges'') connecting a given set of points (``vortices'') without closed loops,
such that the sum of the edge
lengths is a minimum (Kruskal \cite{kruskal56}; Prim \cite{prim57}; Gower \& Ross \cite{gr69}).
In astrophysics, minimum spanning trees have so far mainly been used to analyse the structure of
galaxy clusters (e.g. Barrow et al.\ \cite{bbs85}; Adami \& Mazure \cite{am99};
Doroshkevich et al.\ \cite{doro04}).
From the MST the normalised mean edge length \mbar\ is derived. Unlike the mean
separation length $s$, the mean edge length $m$ depends on the number of stars in the cluster,
therefore it has to be normalised by the factor $(A/n)^{1/2}$
(Marcelpoil \cite{marcelpoil93}),
where $n$ is the total number of stars and $A$ the two-dimensional area of
the cluster.
In the three-dimensional case the normalisation factor is $(V/n)^{1/3}$,
where $V$ is the volume of the cluster.
Area and volume are defined by the normalised convex hull of the objects
(see Appendix~\ref{app:area}).
The normalisation factors are discussed in detail in Appendix~\ref{app:mst}.

An additional reducing operation, called {\em separating}, can be used to isolate subclusters
(Barrow et al.\ \cite{bbs85}). Separating means removing all edges of the MST
whose lengths exceed a certain limit.
When removing edges from a MST, each remaining subgraph is again a MST of its
vortices (Robins et al.\ \cite{robins00}).

\subsection{The combined measure $Q$}

The values \sbar\ and \mbar, on their own, can quantify, but cannot distinguish between, a smooth
large-scale radial density gradient and multiscale fractal subclustering.
Dussert et al.\ (\cite{dussert86}) combine the mean edge length $m$ of a MST and its standard deviation
$\sigma_{\rm m}$ and use the ($m$, $\sigma_{\rm m}$)-plane to separate different
degrees of order in various systems. However, \cite{cw04} show that this is not sufficient to
differentiate between a smooth large-scale radial density gradient and fractal subclustering.
Therefore, \cite{cw04} introduced the parameter $Q = \bar{m}/\bar{s}$, which can provide this distinction.
Large $Q$ values ($Q \ge 0.8$) indicate centrally concentrated clusters having a volume
density $n \propto r^{- \alpha}$, where $Q$ increases with increasing $\alpha$ (i.e. with
increasing degree of central concentration).
Small $Q$ values ($Q \le 0.8$) describe clusters with fractal substructure, where $Q$ decreases with
increasing degree of subclustering.

\subsection{The elongation of a cluster}

We also investigate the elongation of the clusters.
We define the elongation $\xi$ of a cluster as the ratio of the cluster radius
defined by an enclosing circle to the cluster radius derived from the normalised convex
hull of the objects.
A value of $\xi \approx 1$ describes a spherical cluster, while a value of
$\xi \approx 3$ corresponds to an elongated elliptical cluster with an axis ratio of
$a/b \approx 10$.
See Appendix~\ref{app:elongation} for details.

\section{Observations}
\label{sec:obsdata}

 \begin{figure*}
  \centering
  \includegraphics[width=15cm]{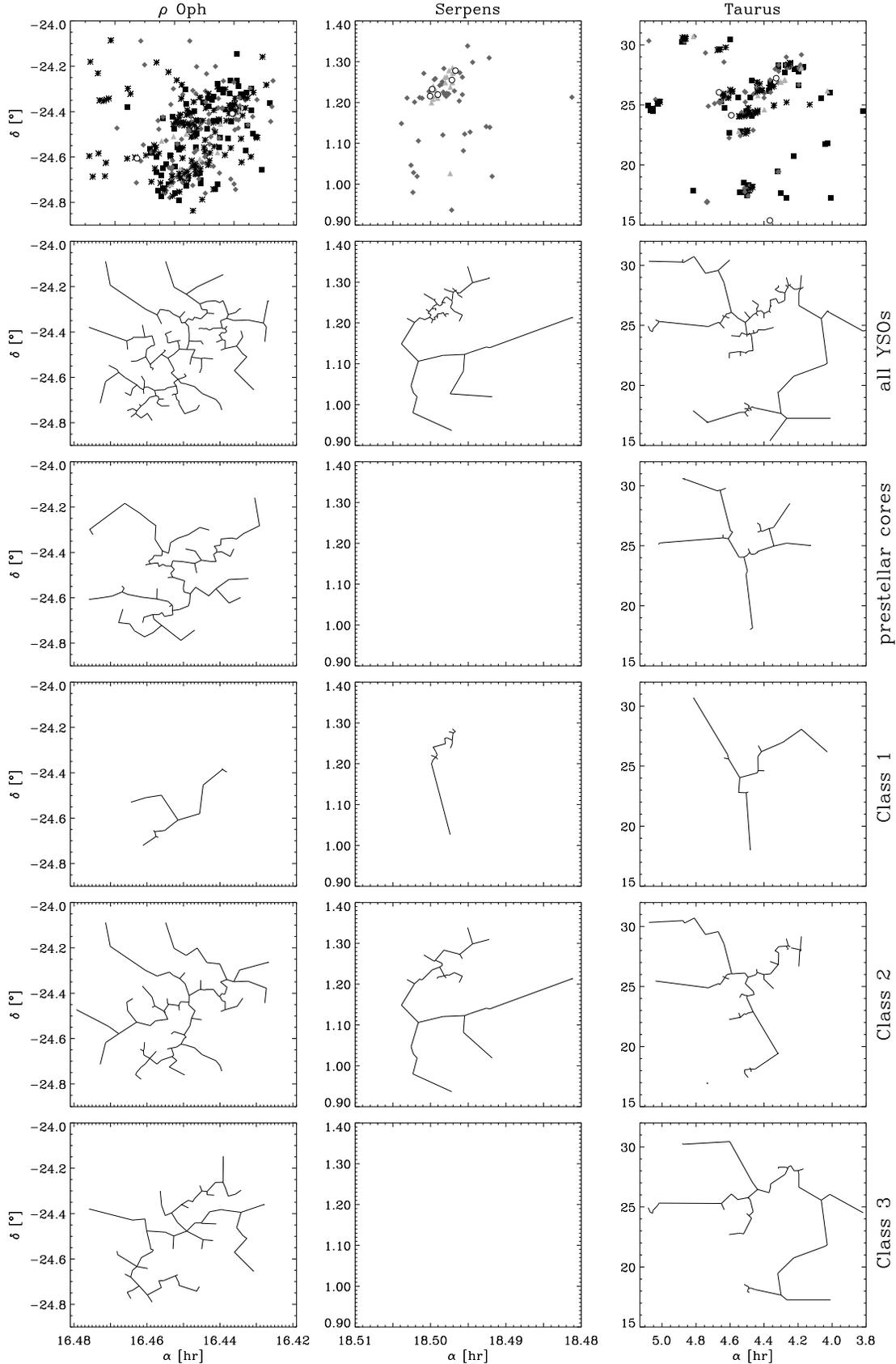}
    \caption{Upper panel:
           Observational data for the star-forming clusters $\rho$~Ophiuchi,
            Serpens, and Taurus. Circles: Class~0, triangles: Class~1, diamonds: Class~2,
	     squares: Class~3, asterisks: prestellar cores.
	     The sources of the observational data are given in the text.
	     Other panels: The MSTs of the same regions for all YSOs (second row), prestellar
	     cores (third row), and Class 1, 2, 3 objects (fourth, fifth, and sixth row).
	     For Serpens, no data on prestellar cores and Class~3 objects are available. }
              \label{fig:obs_pos}
   \end{figure*}

Our observational data are based on the sample of YSOs in embedded clusters compiled from various
sources as discussed in detail by Schmeja et al.\ (2005; hereafter \cite{skf05}).
However, in the current analysis only the regions
$\rho$~Ophiuchi, Taurus and Serpens will be studied in detail. These are clusters,
where sufficient information on the evolutionary classes as well as on the positions
is given in the literature.
IC~348 and Chamaeleon~I are used for determining additional clustering parameters.
The data of $\rho$~Ophiuchi are taken from Bontemps et al.\ (\cite{bontemps01})
(YSOs) and Stanke et al.\ (\cite{stanke04}) (prestellar cores), the Serpens
data are from Kaas et al.\ (\cite{kaas04}) (Class 1/2),
Hurt \& Barsony (\cite{hb96}), and Froebrich (\cite{froebrich05}) (Class~0), the Taurus data are
taken from Hartmann (\cite{hartmann02}) (YSOs) and Lee \& Myers (\cite{lee_myers99}) (prestellar cores).
The data of IC~348 are taken from Luhman et al.\ (\cite{luhman03}) and those of Chamaeleon~I
from C\'ambresy et al.\ (\cite{cambresy98}).
For further details on the compilation of the original sample see \cite{skf05}.
The numbers of objects are given in Table~\ref{tab:obs_results}, the positions of all
YSOs are plotted in the upper panel of Fig.~\ref{fig:obs_pos}.
The adopted distances used to determine the absolute values of \sbar\ and \mbar\ are 140~pc
for $\rho$~Oph (Bontemps et al.\ \cite{bontemps01}), 260~pc for Serpens
(Kaas et al.\ \cite{kaas04}), 140~pc for Taurus (Hartmann \cite{hartmann02}),
315~pc for IC~348 (Luhman et al.\ \cite{luhman03}), and 150~pc for Cha~I
(Haikala et al.\ \cite{haikala05}).

\section{The models}
\label{sec:models}

\begin{figure*}
  \centering
  \includegraphics[width=17cm]{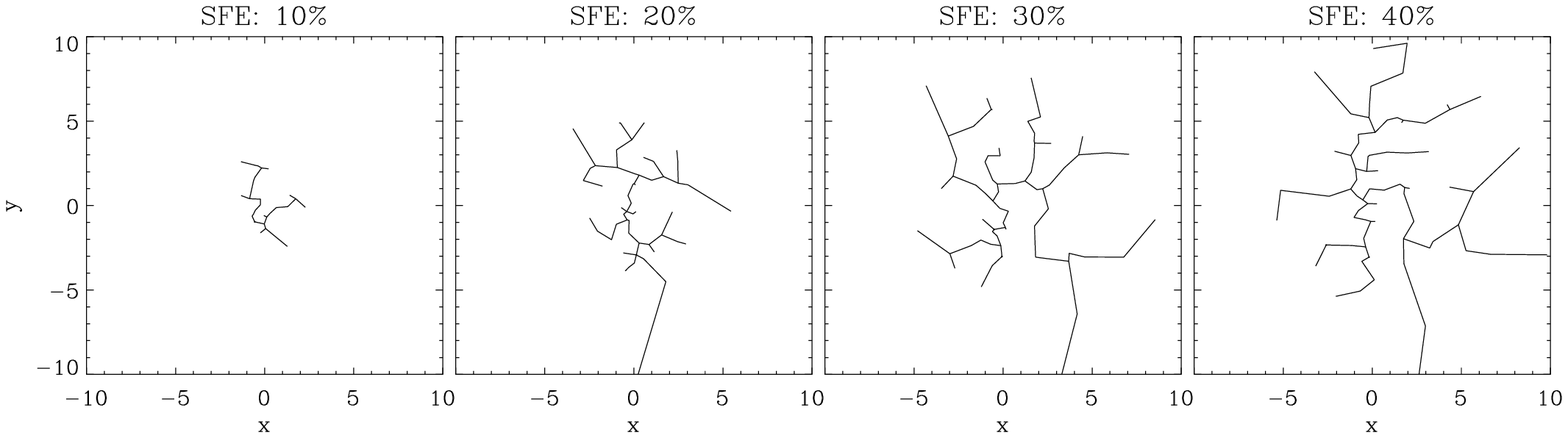}
    \caption{The 2D minimum spanning tree for the YSOs of model M6k4a projected into the xy-plane
    at a star formation efficiency of 10\%, 20\%, 30\% and 40\% (from left to right).
	 }
              \label{fig:model_mst}
   \end{figure*}

We perform numerical simulations of the fragmentation and collapse of
turbulent, self-gravitating gas clouds and the resulting formation and
evolution of protostars as described in Schmeja \& Klessen
(2004; hereafter \cite{sk04}).  We use a code based on smoothed particle hydrodynamics
(SPH; Monaghan \cite{monaghan92}) in order to resolve large density
contrasts and to follow the evolution over a long timescale.  The code
includes periodic boundary conditions (Klessen \cite{klessen97}) and
sink particles (Bate et al.\ \cite{bbp95}) that replace high-density
cores while keeping track of mass and linear and angular momentum.
We determine the resolution limit of our SPH calculations using the Bate
\& Burkert (\cite{bate_burkert97}) criterion. This is sufficient for the
highly nonlinear density fluctuations created by supersonic turbulence as
confirmed by convergence studies with up to $10^7$ SPH particles
(Jappsen et al.\ \cite{jappsen05}; Li et al.\ \cite{li05}).

Our simulations consist of two globally unstable models that contract
from Gaussian initial conditions without turbulence and of 22 models
where turbulence is maintained with constant rms Mach numbers $\cal
M$, in the range $0.1 \le {\cal M} \le 10$.  We distinguish between
turbulence that carries its energy mostly on large scales, at
wavenumbers $1 \le k \le 2$, on intermediate scales, i.e.\ $3 \le k
\le 4$, and on small scales with $7 \le k \le 8$.  The naming of the
models, G1 and G2 for the Gaussian runs, and M\mach k$k$ (with rms
Mach number \mach\ and wavenumber $k$) for the turbulent models,
follows \cite{sk04}.  Details of the individual
models are given in their Table~1.

The dynamical behaviour of isothermal self-gravitating gas is scale free
and depends only on the ratio $\alpha$ between internal energy and potential energy:
$\alpha = E_{\rm int}/|E_{\rm pot}|$.
Since we are only interested in the positions of the young stars, and since
all the clustering parameters are normalised by the cluster radius, the physical
scaling is irrelevant for the present study.

The YSO classes are determined as follows (see \cite{skf05} and Froebrich et al.\ \cite{fssk05}
for a detailed discussion): The beginning of Class~0 is
identified with the formation of the first hydrostatic core, when
the central object has a mass of about $0.01~\mathrm{M}_{\sun}$
(Larson \cite{larson03}).  The transition from Class~0 to Class~1 is
reached when the envelope mass is equal to the mass of the central
protostar (Andr\'e et al.\ \cite{andre00}).
We determine the transition from Class~1 to Class~2 when the optical
depth of the remaining envelope becomes unity at $2.2~\mu \mathrm{m}$ ($K$-band).
The end of Class~0 and Class~1 corresponds to a mass of about 0.43
and 0.85 times the final mass, respectively (\cite{skf05}).
Lacking a feasible criterion to distinguish Class~2 from Class~3 objects,
we consider both classes combined.
Prestellar cores are identified by a clump-finding algorithm described in
Klessen \& Burkert (\cite{klessen_burkert00}; see also \cite{skf05}).

We only consider models with a numerical resolution of at least 200\,000 particles.
Furthermore, in order to get reasonable numbers of protostars for the
statistics, we select only those models where more than 37 protostars with
$M_\mathrm{end} \ge 0.1~\mathrm{M}_{\sun}$
(roughly corresponds to the detection limits of the observations)
are formed.
This reduces our set of models to 16. Again, see Table 1 of \cite{sk04}
for further details.

\section{Application to the data}
\label{sec:application}

We construct the MST following Prim's (\cite{prim57}) algorithm, as also described by
Gower \& Ross (\cite{gr69}).
For both the observations and the models, the parameters \sbar, \mbar, $Q$, and $\xi$ are computed for
the different evolutionary classes independently (provided, there is a sufficiently large number of
YSOs of that class) as well as for the entire cluster.
In no region is the number of Class~0 sources large enough to be included.
In addition, these parameters are determined at
frequent timesteps of the simulations to obtain the temporal evolution of the parameters.
In the case of \mbar, this requires to construct the MST anew at every chosen timestep.
As an example, Fig.~\ref{fig:model_mst} shows the MST of one model at different evolutionary stages.
While the parameters \sbar, \mbar\ and $Q$ are calculated using the normalisation factor given above,
$\bar{s}^*$, $\bar{m}^*$ and $Q^*$ are determined with the normalisation factor and cluster
radius of \cite{cw04} (see also the discussion in the Appendix).

As shown by \cite{cw04}, the effect of binary stars on the clustering parameters
is not negligible.
Since binaries create very short edge lengths, a large fraction of binaries will
significantly reduce the mean edge length \mbar\ (and as a consequence also change $Q$).
As the binaries are not part of the clustering regime, their influence
on the clustering parameters has to be minimized.
While for most of the clusters it is not relevant, Taurus is known to
have a large binary population.
Thus we removed the known binary companions from our sample of YSOs in Taurus.

We calculate the clustering parameters \sbar, \mbar, $Q$, and $\xi$ for the
model cluster in three dimensions, and, in order to compare them with the
apparent two-dimensional observational data, projected into the
xy-, xz-, and yz-plane.
The positions of the stars are corrected for the periodic boundary conditions.
Only objects inside a box of ten times the
side length of the original computational box are considered,
if there are any objects outside this volume, we assume that they
have left the cluster and are not relevant for the clustering process.

In the beginning stages of the cluster as a whole, and of the individual evolutionary
classes, the number of objects is rather low.
In such cases the analysis may not be statistically significant and
has to be taken with a grain of salt.

  \begin{figure*}
   \centering
  \includegraphics[width=17cm]{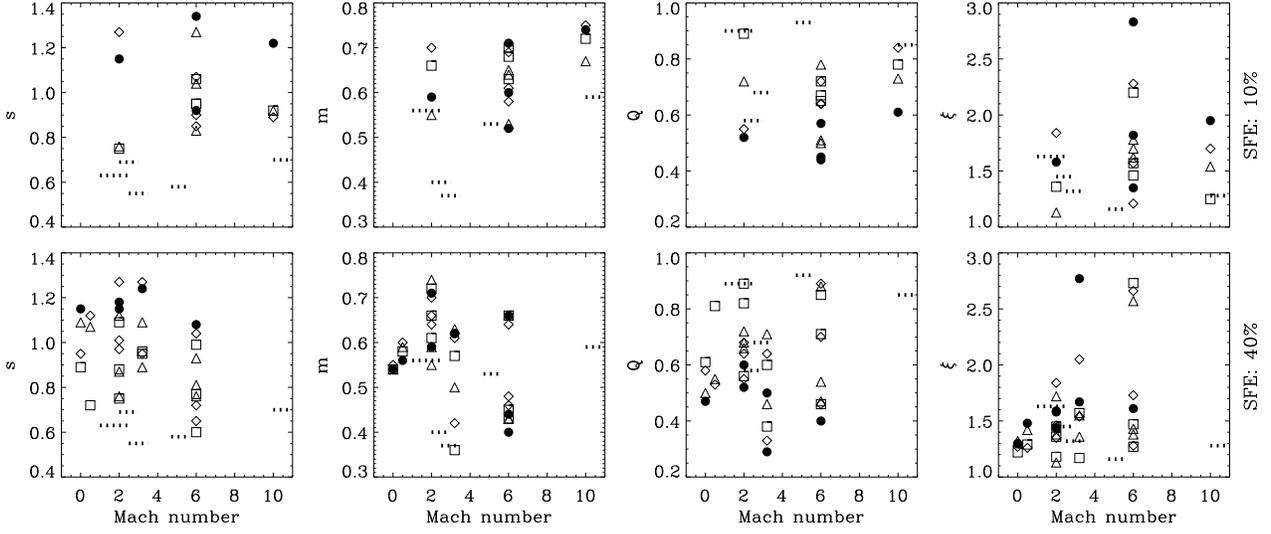}
      \caption{The clustering parameters \sbar, \mbar, $Q$, and $\xi$ for all models at
      a SFE of 10\% (upper panel) and 40\% (lower panel), plotted versus the Mach number.
      Shown are the values for the projection into the xy-, xz-,
      and yz-plane (diamonds, triangles and squares, respectively), and for the 3D analysis
      (filled circles).
      The horizontal lines show the corresponding values from the observational data for
      Serpens, Taurus, Chamaeleon~I, IC~348, and $\rho$~Oph (from left to right).
              }
         \label{fig:parameters}
 \end{figure*}

\section{Discussion}
\label{sec:discussion}

\subsection{Observations}
\label{sec:discuss_obs}

Table~\ref{tab:obs_results} lists the clustering measures introduced in Sect.~\ref{sec:statistics} for the observations.
Columns 4 to 7 list the (non-normalised) mean separation of objects and mean MST edge lengths
in arc\-minutes and parsec. Columns 8 to 10 give the parameters \sbar, \mbar\ and $Q$, while
Columns 11 to 13 list the same values calculated using the normalisation of \cite{cw04}
($\bar{s}^*$, $\bar{m}^*$, $Q^*$).
The last column gives the elongation $\xi$.
The values of $\bar{s}^*$, $\bar{m}^*$ and $Q^*$ agree with the values given by \cite{cw04}, except for $\bar{s}^*$ and $\bar{m}^*$ of Taurus
(interestingly, however, our $Q^*$ is the same as theirs).
We attribute small differences to the slightly different underlying samples, and
the discrepancy for Taurus to the different treatment of binaries.
Due to the different definition of the radius/area of the cluster, $\bar{s}^*$ and
$\bar{m}^*$ differ significantly from \sbar\ and \mbar, while $Q$ and $Q^*$ are roughly
the same, in particular for large samples (see the discussion in Appendix~\ref{app:mst}).
Taurus and Chamaeleon~I have substructure, while $\rho$~Oph and
IC~348 are centrally concentrated clusters (see the discussion in \cite{cw04}).
Serpens (not discussed by \cite{cw04}) has $Q = 0.90$, corresponding to a central
concentration with a radial density exponent $\alpha \approx 1.9$, similar
to IC~348.

The linear distances $s$ and $m$ are in the same range for $\rho$~Oph,
Serpens, and IC~348, significantly larger for Chamaeleon~I, and about an order
of magnitude larger in the case of Taurus, confirming the notion that Taurus
represents a somewhat less clustered mode of star formation.
However, when considering only the central part of the Taurus region, the
values decrease significantly to $s = 3.05$\,pc and $m = 0.49$\,pc.
The latter is comparable to the 0.3\,pc estimated as the average distance
to the nearest stellar neighbour in the central region of Taurus by Gomez
et al.\ (\cite{gomez93}) and Hartmann (\cite{hartmann02}).

The mean separation $s$ increases with the evolutionary class
for all three clusters investigated in this regard, reflecting
the expansion of the cluster, which can also be seen in Fig.~\ref{fig:obs_pos}.
While Class~0 protostars are formed in the high-density central regions, more
evolved YSOs already had time to move to more remote regions (see also Kaas et al.\
\cite{kaas04}).
Note that the prestellar cores do not fit into this sequence, as they are
distributed over an area roughly as large as the entire cluster.
Thus we speculate that not all objects identified as prestellar cores will eventually
form stars.
Only in the central parts of the cluster may the density be high enough to make the
cores collapse.
This is consistent with the findings of V{\'a}zquez-Semadeni et al.\ (\cite{vksb05})
that a significant number of ``failed cores'' should exist, which may redisperse
and which may correspond to the observed starless cores.

The elongation of the clusters ranges from $\xi = 1.16$ for the almost perfectly
spherical cluster IC~348 to $\xi = 1.63$ for Serpens. The $\xi$ value can
differ significantly for subclusters, e.g. the filamentary central part of the
Taurus region has an elongation of $\xi = 1.83$.

\subsection{Models}

 We compute the clustering parameters \sbar, \mbar, $Q$, and $\xi$ for all models and compare
 them with each other as well as with those from the observations.
 Figure~\ref{fig:parameters} shows the clustering parameters for all models, sorted by
 the Mach number at a star formation efficiency (SFE) of 10\%  and 40\%.
 The clustering parameters of the observed clusters (from Table~\ref{tab:obs_results})
 are shown as horizontal lines at their Mach number (taken from
 \cite{skf05}).
The \sbar\ and \mbar\ values of the models are in general significantly larger than
those from the observations.
A particularly large discrepancy is noted when the model cluster is strongly elongated.
A large $\xi$ value reduces the normalisation factor and increases \sbar\ and \mbar.
While the observed clusters show rather moderate elongations in the range
$1.1 < \xi < 1.6$, in some models the stars form in a single filament with an elongation
$\xi > 3$.
Model clusters with elongations in the range of the observations show
also a good agreement in the other parameters. For example, the fairly spherical
cluster of model M6k4a (shown in Fig.~\ref{fig:model_mst}) has (at 40\% SFE and
projected into the xy-plane) an elongation of $\xi = 1.28$ and the clustering
parameters $\bar{s} = 0.72$, $\bar{m} = 0.64$, and $Q = 0.89$. These values
are almost identical to those of $\rho$~Ophiuchi ($\xi = 1.28$,
$\bar{s} = 0.70$, $\bar{m} = 0.59$, $Q = 0.85$).
For the other projections the values differ by less than 8\%.
Most of the $Q$ values, which are independent of the area, lie in the
same range as those from the observations.
We find no correlation of the clustering parameters with the properties of the turbulent
driving (Mach number or wave number $k$) of the models.
Neither do the values from the observations show any correlation with the Mach number. Thus we conclude
that if there is any systematic influence of the turbulent environment on the clustering behaviour,
it is only existent in the earliest phase of cluster formation, before it is smoothed out
by the motions of the individual protostars (see also Bate et al.\ \cite{bcm98}).

 \begin{figure*}
  \centering
  \includegraphics[width=17cm]{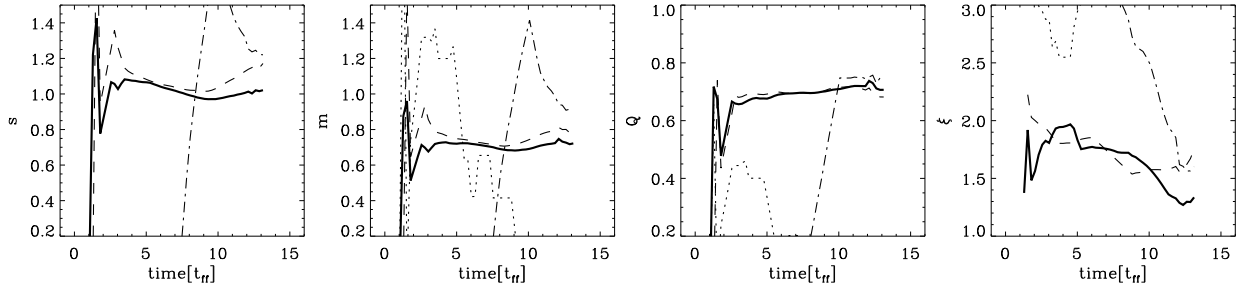}
    \caption{The temporal evolution of the 3D clustering parameters $\bar{s}$, \mbar,
      $Q$, and $\xi$ for one model (M6k4a), dotted: Class~0, dashed: Class~1, dash-dotted: Class~2/3,
      solid line: entire cluster.}
              \label{fig:sequence}
   \end{figure*}

    \begin{figure*}
  \centering
  \includegraphics[width=17cm]{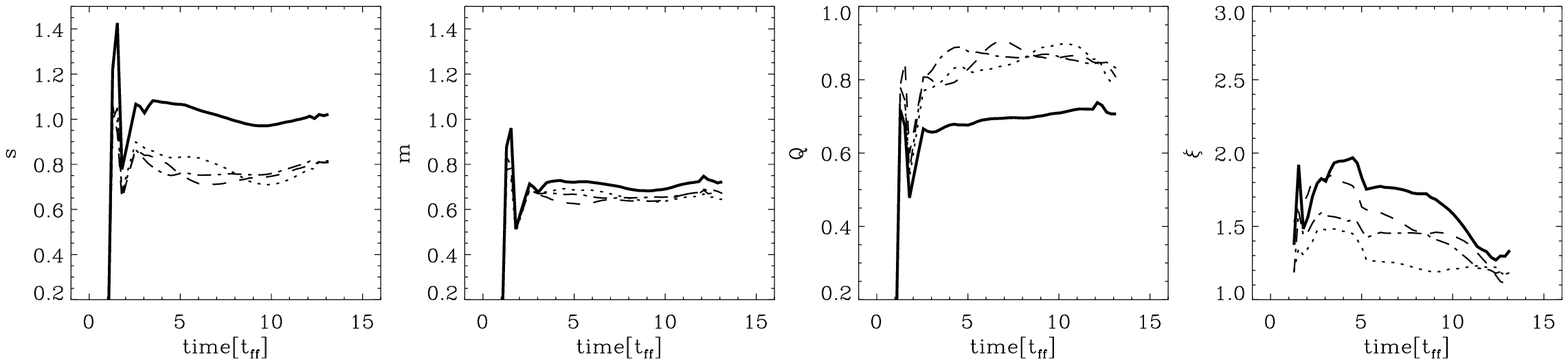}
    \caption{The temporal evolution of the clustering parameters $\bar{s}$, \mbar,
      $Q$, and $\xi$ of the model M6k4a for the 3D analysis (solid line) and for the
      projection into the xy (dotted), xz (dashed), and yz plane (dash-dotted).
       }
              \label{fig:sequence_xyz}
   \end{figure*}

Fig.~\ref{fig:model_mst} shows the MST of the model cluster M6k4a at different
evolutionary stages and reveals the expansion of the cluster.
We analyse the temporal evolution of the clustering parameters: \sbar$(t)$, \mbar$(t)$, $Q(t)$, $\xi(t)$
in all models. As an example, Fig.~\ref{fig:sequence} shows this sequence for the said model M6k4a.
The general behaviour is the same for all models:
\sbar\ and \mbar\ decline slightly with time, while $Q$ increases slowly or stays at a roughly
constant value. This evolution is shown by the entire cluster as well as by the individual
classes, although in later stages the latter values might fall to zero as the number of
objects in a particular class becomes zero.
Decreasing \sbar\ and \mbar\ values indicate that star formation sets in in different,
rather dispersed regions of the cloud. The cluster becomes denser as more and more gas
is turned into protostars.
New stars are formed faster than the cluster expands.
According to the $Q$ values, the cluster evolves slowly from fractal subclustering
to a more centrally concentrated cluster
although in no model does $Q_{2D}$ rise significantly above the `divding value' of 0.8.
This again shows that the cluster builds up from separate groups which will grow into
a single, more centrally concentrated cluster, as also found by Bonnell et al.\
(\cite{bonnell03}) and Clark et al.\ (\cite{clark05}).

\subsection{The effect of projection}

\label{sec:projection}
 \begin{figure*}
  \centering
  \includegraphics[width=17cm]{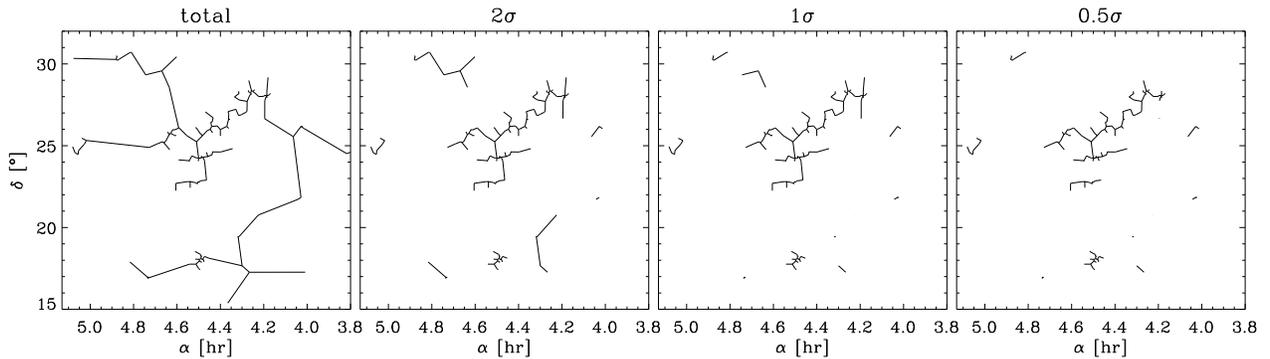}
    \caption{The MST of the Taurus cluster as a whole (left panel) and separated at
     $l+ 2 \sigma_\mathrm{m}$, $l+ \sigma_\mathrm{m}$,
     and $l+ 0.5 \sigma_\mathrm{m}$ (from left to right).
       }
              \label{fig:mst_separated}
   \end{figure*}

 \begin{figure*}
  \centering
  \includegraphics[width=17cm]{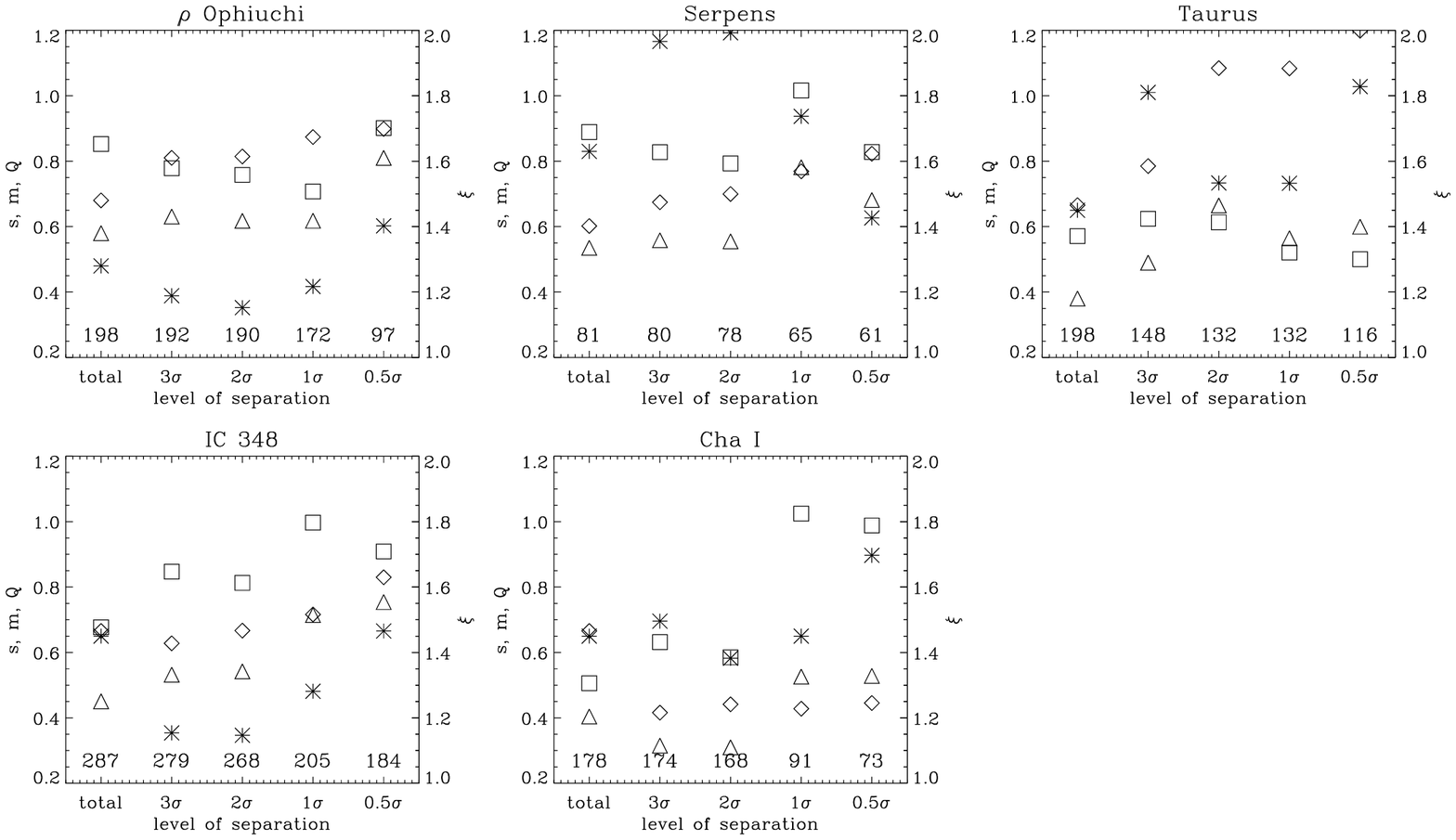}
    \caption{The clustering parameters \sbar\ (diamonds), \mbar\ (triangles),
       $Q$ (squares), and $\xi$ (asterisks; scale on the right-hand axis)
       of the largest remaining subcluster at different levels of separation.
       The numbers along the abscissa give the number of YSOs contained in the
       particular subcluster.
       }
              \label{fig:smQ_separated}
   \end{figure*}

Looking at the two-dimensional projections of the 3D model clusters does not significantly
change the picture as a whole. The individual \sbar, \mbar\ and $Q$ values can
indeed differ for the projection into the xy, xz, and yz plane, but the
qualitative behaviour of the evolution is more or less the same, independent of
the projection (Fig.~\ref{fig:sequence_xyz}).
While the \sbar$_\mathrm{3D}$ and \mbar$_\mathrm{3D}$ values usually are higher than the values of the
projections, $Q_\mathrm{3D}$ tends to be lower than the values of the projections.
The investigation of several hundred randomly created clusters shows that
$\bar{s}_\mathrm{3D}$ is always expected to be larger than the value for the projections
$(\bar{s}_\mathrm{3D}/\bar{s}_\mathrm{2D} \approx 1.2)$, while
the \mbar\ values can be in the range
$0.95 \lesssim \bar{m}_\mathrm{3D}/\bar{m}_\mathrm{2D} \lesssim 1.4$ with a mean
value of 1.1.
In the extreme case, $Q_\mathrm{3D}$ can differ by up to $30\%$ from the 2D value.
Note that the physical interpretation of $Q$ as given by \cite{cw04} is based
on the two-dimensional analysis. Therefore, for an interpretation of the numerical
values (and not only the trend) of $Q$, the projected values have to be used.

The elongation measure $\xi$, on the other hand, depends strongly on the projection.
Obviously, an elongated, filament-like structure seen from the side
will look spherical when observed along its major axis.
In the case of the models the three-dimensional value of $\xi$ can be used to
describe the true shape of the cluster.
A large scatter in the $\xi_\mathrm{2D}$ values means a high $\xi_\mathrm{3D}$ value
and vice versa.

\subsection{The effect of separating}

We separate the observed clusters by succesively removing all MST edges with lengths
$l$ larger than $l+ 3 \sigma_\mathrm{m}$, $l+ 2 \sigma_\mathrm{m}$, $l+ \sigma_\mathrm{m}$,
and $l+ 0.5 \sigma_\mathrm{m}$, where $\sigma_\mathrm{m}$ denotes the standard deviation
of the mean edge length of the MST.
This process is demonstrated for the Taurus star-forming region in
Fig.~\ref{fig:mst_separated}.
When we compare all three regions with sufficient data, we see that for
Serpens and Taurus only solitary stars in the outskirts of the
cluster are removed, while the central part of the cluster stays connected.
However, the more homogeneous $\rho$~Oph cluster
breaks down in two roughly equally large clusters at the last step.
In Fig.~\ref{fig:smQ_separated} we show the clustering parameters of the largest
remaining subcluster (i.e., the one containing the largest number of stars after
each level of separation).
The parameter $Q$ is less affected by the separating procedure than \sbar\ and \mbar.
Significant changes are only seen when large parts of the cluster are excluded
(as in the step from 1 to $0.5 \sigma$ for $\rho$~Oph, when the cluster breaks up
into two large subclusters).
The elongation measure $\xi$ varies significantly with the level of separation.
For example, the remaining subcluster in Taurus shows a filamentary structure
and thus a larger $\xi$ ($\xi = 1.83$) than the cluster as a whole.

\section{Summary and Conclusions}
\label{sec:conclude}

We show that the normalised mean correlation length \sbar\ and the mean edge length
of the minimum spanning tree \mbar, and in particular the combination of both
parameters, $Q$, as proposed by \cite{cw04},
are very useful tools to study the structures of star clusters, both observed
ones and those from numerical simulations.
We refine the definiton of the cluster area by using the normalised convex hull
rather than a circular or rectangular area around the objects.
Unlike \sbar\ and \mbar\ the parameter $Q$ is independent of the definiton of the area
of the cluster. In addition, it is less affected when removing stars from the cluster
by separating the MST.

We introduce a new measure $\xi$ for the elongation of a cluster. It is
defined as the ratio of the cluster radius determined by an enclosing
circle to the cluster radius derived from the normalised convex
hull. This is a stable statistical measure not influenced by fractal
substructre, which could also be applied to the filamentary structure of
molecular clouds.

The mean separation $s$ increases with the evolutionary class, reflecting
the expansion of the cluster. The prestellar cores do not follow that sequence,
leading us to the speculation that not all objects classified as prestellar
cores will eventually form stars.
The clustering values of the models lie roughly in the same range as those
from observed clusters.
A particularly good agreement is reached when the clusters have similar elongation
values $\xi$.
No correlation with the Mach number or the wave number of the turbulence
is found. We conclude that possible influences of the turbulent environment
on the clustering behaviour are quickly smoothed out by the velocity dispersion
of the young stars.

The temporal evolution of the clustering parameters shows that the star cluster
builds up from several subclusters and evolves to a more centrally concentrated
state.
New stars are formed faster than the cluster expands.
The projection of the 3D models into a 2D plane changes the clustering parameters,
but not the general behaviour with time.

\begin{acknowledgements}
This work is funded by the Emmy Noether Programme of the {\em Deutsche
Forschungsgemeinschaft} (grant no.\ KL1358/1).
We are very grateful to Roland Gredel, Thomas Stanke, Michael Smith
and Tigran Khanzadyan for providing us with their $\rho$~Oph data
prior to publication and to Lee Hartmann for sending us his data of
the Taurus cloud.
We wish to thank Dan Kushnir and Spyridon Kitsionas for valuable discussions
and the referee, Anthony Whitworth, for his prompt report.
S.\,S. acknowledges the hospitality of the Institute for Pure and Applied Mathematics,
University of California, Los Angeles, during part of this work.

\end{acknowledgements}


\begin{appendix}

\section{Radius and area of a cluster}
\label{app:area}

For the normalisation of the parameters \sbar\ and \mbar\ the radius and the
area (or volume) of the cluster are needed (see Appendix~\ref{app:mst}).
Since a star cluster has no well-defined natural boundary, different
approaches to determine the cluster area have been used.
\cite{cw04} define the cluster radius as the distance
between the mean position of all cluster members and the most distant star and
the area as a circle with the cluster radius: $A = \pi R_{\rm cluster}^2$.
Adami \& Mazure (\cite{am99}) define the area used for the normalisation of the MST
edge lengths as the maximum rectangle of the point set.
However, both methods tend to significantly overestimate the area of the cluster, in particular
if the cluster is elongated or irregularly shaped rather than spherical.
Therefore, we estimate the area $A$ of the cluster using the convex hull
of the data points, normalised by an additional geometrical factor taking
into account the ratio of the number of objects inside and on the convex hull:
\begin{equation}
A = m(H)/[1-(v_\mathrm{H}/n)]
\label{eq:norm_area}
\end{equation}
(Hoffman \& Jain \cite{hj83}; Ripley \& Rasson \cite{rr77}),
where $H$ is the convex hull, $m(H)$ its area, $v_\mathrm{H}$ the number of vertices on $H$
and $n$ the number of objects.
The correction factor is used since the convex hull by itself tends to be smaller
than the true sampling window.
To be consistent, we define the cluster radius as the radius of a circle with the
same area $A$.
In the three-dimensional case the volume $V$ and the radius are defined analogously.
The convex hull is computed using the programme {\em Qhull}
(Barber et al.\ \cite{barber96}).

 \begin{figure}
  \centering
    \resizebox{\hsize}{!}{\includegraphics{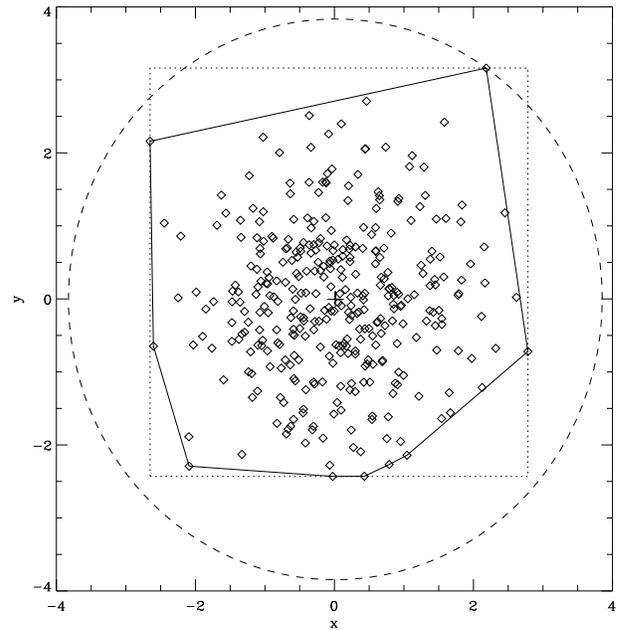}}
      \caption{Randomly distributed data points and the area they span
      according to different definitions (circle, rectangle, convex hull).
              }
         \label{fig:cluster_area}
 \end{figure}

Fig.~\ref{fig:cluster_area} demonstrates that the definition of the cluster area
is crucial, since it can differ by a factor of two or more. In the given example
of randomly distributed data points,
the area of the circle is 46.33, that of the rectangle is 30.40, the
area of the convex hull is 24.19, and the normalised area (Eq.~\ref{eq:norm_area})
is 24.83.
For the calculation of the parameter $Q = \bar{m}/\bar{s}$ the size
of the radius/area is irrelevant (as long as the area used for the normalisation of
\mbar\ depends on the radius used for the normalisation of \sbar\ or vice versa).
The radius is cancelled since it is used to normalise both, \mbar\ and \sbar.

\section{Normalisation of the Minimum Spanning Tree}
\label{app:mst}

\begin{figure}
  \centering
    \resizebox{\hsize}{!}{\includegraphics{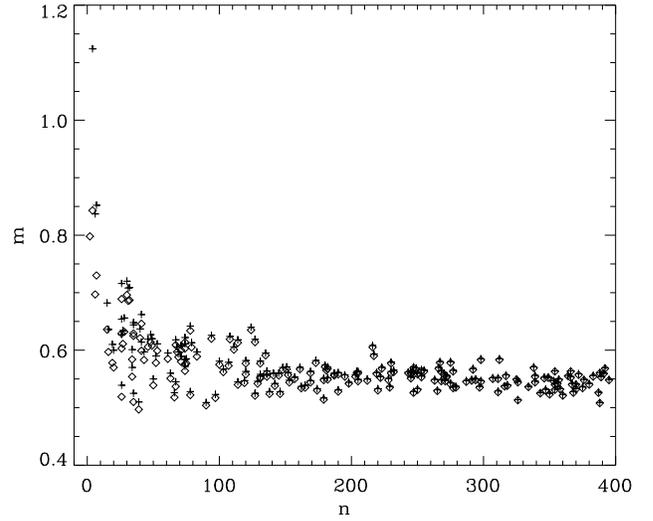}}
      \caption{Mean edge lengths \mbar\ for 200 randomly created 2D sets of points ($3 \le n \le 400$),
             normalised with the factor~\ref{eq:norm_m93} (crosses) and \ref{eq:norm_cw04}
	     (diamonds), respectively,
	     plotted versus the number of points.
              }
         \label{fig:mst_normalisation}
 \end{figure}

The edge lengths of a minimum spanning tree depend on the number of points and on the area,
therefore the mean edge length $m$ has to be normalised, in order to compare the results
from samples with different numbers of objects and/or different areas.
Beardwood et al.\ (\cite{bhh59}) show that the length of the shortest closed path through $n$ points
$l(P^n)$ in a plane region of area $A$ is asymptotically proportional to $\sqrt{n A}$ for large $n$.
Since the number of edges in a MST is $(n-1)$, Hoffman \& Jain (\cite{hj83}) and \cite{cw04}
claim that the expected length of a randomly selected edge of a MST is asymptotically proportional
to
\begin{equation}
\label{eq:norm_cw04}
\frac{\sqrt{n A}}{n-1} .
\end{equation}
 They use this factor to normalise the mean edge lengths of different data
sets. Beardwood et al.\ (\cite{bhh59}), however, deal with the {\em closed path} through $n$ points
(i.e., the travelling salesman problem), not a MST, as cited incorrectly by Hoffman \& Jain (\cite{hj83}) and \cite{cw04}.
Marcelpoil (\cite{marcelpoil93}), on the other hand, deduces the normalisation factor
\begin{equation}
\label{eq:norm_m93}
\sqrt{A/n}
\end{equation}
for comparing the mean edge lengths of different samples.
Since
\begin{equation}
\lim_{n \rightarrow \infty} \frac{\sqrt{n A}/(n-1)}{\sqrt{A/n}} =
\lim_{n \rightarrow \infty} \frac{n}{n-1} = 1,
\label{eq:lim}
\end{equation}
both methods lead to
the same result for large $n$, e.g.\ for $n > 100$ the difference is $\lesssim 1\%$
(see Fig.~\ref{fig:mst_normalisation}).
In this work we use the normalisation factor of Marcelpoil (\cite{marcelpoil93}).
which seems more plausible to us.

In the three-dimensional case the general result for $k$ dimensions of Beardwood et al.\
(\cite{bhh59}) (their Equation~2) can be written as $l(P^n) \propto \sqrt[3]{V n^2}$,
leading to the factor
\begin{equation}
\frac{\sqrt[3]{V n^2}}{n-1}
\end{equation}
to normalise the mean edge lengths of the 3D MST.
However, we follow the reasoning of Marcelpoil (\cite{marcelpoil93})
and use the normalisation factor
\begin{equation}
\sqrt[3]{V/n}
\end{equation}
in the 3D case.
The volume V is the volume of the convex hull as discussed above.
Again, the ratio of both normalisation factors asymptotically approaches 1, making
them interchangeable for large $n$.

Note that we use different definitions of the radius and the area from \cite{cw04}
(Appendix~\ref{app:area}), resulting in an additional difference between our
and their parameters \mbar\ and \sbar.
However, in the calculation of $Q$ the radius contained in the normalisation factor
is cancelled out, so the difference in the $Q$ parameters boils down to
Equation~\ref{eq:lim}.
Unlike \mbar\ and \sbar, the more relevant parameter $Q$
is the same for a large number of objects, independent of the normalisation method.
To allow comparison, we list the parameters computed using the normalisation factor
and the cluster radius according to \cite{cw04} in Table~\ref{tab:obs_results} as well.

\section{Elongation of a cluster}
\label{app:elongation}

 \begin{figure}
  \centering
    \resizebox{\hsize}{!}{\includegraphics{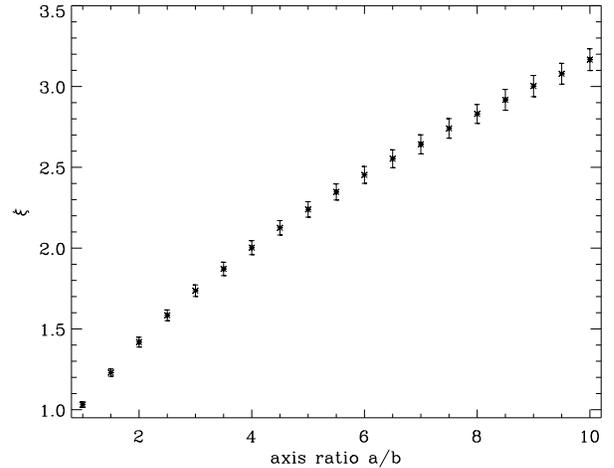}}
      \caption{The relation between the axis ratio $a/b$ of an elliptical
       area and the elongation $\xi$, with the $1 \sigma$ error also indicated.
              }
         \label{fig:elongation}
 \end{figure}

We notice that some of the model clusters are strongly elongated, causing a large difference
in the cluster area depending on whether it is defined by the enclosing circle or by
the normalised convex hull.
We use this fact to propose a new, statistically stable characterisation of the
elongation of a cluster.
We define the elongation measure $\xi$ as the ratio
\begin{equation}
\xi = \frac{R_\mathrm{cluster}^\mathrm{circle}}{R_\mathrm{cluster}^\mathrm{conv.\ hull}}
\end{equation}
This is an approach similar to the ``filament index'' $\mathcal{F}$ introduced by
Adams \& Wiseman (\cite{adams+wiseman94}).
However, while Adams \& Wiseman (\cite{adams+wiseman94}) use the actual area of
a cloud obtained from column density maps, we have to estimate the area of
the cluster of point sources by the convex hull.
This makes $\xi$ a reliable measure for the elongation, since it excludes the possibility
that fractal substructure in an otherwise spherical cluster leads to a large ratio
of the two radii.
The measure $\xi$ could also be used to quantify the filamentary structure in
molecular clouds.

In order to test if $\xi$ is indeed a good measure for the elongation we place 350
points randomly on an elliptical area with increasing axis ratio ($1 \le a/b \le 10$).
To minimise the statistical scatter we perform 500 different realisations for each
$a/b$ and determine the mean $\xi$ and its standard deviation for each step.
Figure~\ref{fig:elongation} shows that $\xi$ increases with the axis ratio $a/b$.
For $a/b = 1$ (i.e., a circle), $\xi \approx 1$, as expected.

\end{appendix}


\begin{thebibliography}{}



\bibitem[1999]{am99} Adami, C., \& Mazure, A. 1999, \aaps, 134, 393

\bibitem[1994]{adams+wiseman94} Adams, F.~C., \& Wiseman, J.~J. 1994, \apj, 435, 693

\bibitem[2000]{andre00} Andr\'{e}, P., Ward-Thompson, D., \& Barsony, M. 2000, in
      Proto\-stars and Planets IV, ed.\ V. Mannings, A.~P. Boss, \& S.~S. Russell
      (Tucson: University of Arizona Press), 59

\bibitem[1996]{barber96} Barber, C.~B., Dobkin, D.~P., \& Huhdanpaa, H.~T. 1996,
ACM Trans\-actions on Mathematical Software, 22, 469, http://www.qhull.org

\bibitem[1985]{bbs85} Barrow, J.~D., Bhavsar, S.~P., \& Sonoda, D.~H. 1985, \mnras, 216, 17

\bibitem[1997]{bate_burkert97} Bate, M.\ R., \& Burkert, A. 1997, \mnras, 288, 1060

\bibitem[1995]{bbp95} Bate, M.~R., Bonnell, I.~A., \& Price, N.~M. 1995, \mnras, 277, 362

\bibitem[1998]{bcm98} Bate, M.~R., Clarke, C.~J., \& McCaughrean, M.~J. 1998, \mnras, 297, 1163

\bibitem[1959]{bhh59} Beardwood, J., Halton, J.~H., \& Hammersley, J.~M. 1959,
Proc.\ Cambridge Philos.\ Soc., 55, 299

\bibitem[2003]{bonnell03} Bonnell, I.~A., Bate, M.~R., \& Vine, S.~G. 2003, \mnras, 343, 413

\bibitem[2001]{bontemps01} Bontemps, S., Andr\'e, P., Kaas, A.~A., et al.\ 2001,
       \aap, 372, 173

\bibitem[1998]{cambresy98} Cambr\'esy, L., Copet, E., Epchtein, N., et al.\ 1998, \aap, 338, 977

\bibitem[CW04]{cw04} Cartwright, A., \& Whitworth, A.~P. 2004, \mnras, 348, 589 (CW04)

\bibitem[2005]{clark05} Clark, P.~C., Bonnell, I.~A., Zinnecker, H., \& Bate, M.~R.
        2005, \mnras, 359, 809

\bibitem[2004]{doro04} Doroshkevich, A., Tucker, D.~L., Allam, S., \& Way, M.~J. 2004, A\&A, 418, 7

\bibitem[1986]{dussert86} Dussert, C., Rasigni, G., Rasigni, M., Palmari, J., \& Llebaria, A. 1986,
Phys.\ Rev.\ B, 34, 3528

\bibitem[2005]{froebrich05} Froebrich, D. 2005, \apjs, 156, 169

\bibitem[2005]{fssk05} Froebrich, D., Schmeja, S., Smith, M.~D., \& Klessen, R.~S. 2005, MNRAS, submitted

\bibitem[1999]{gladwin99} Gladwin, P.~P., Kitsionas, S., Boffin, H.~M.~J., \& Whitworth, A.~P.
      1999, \mnras, 302, 305

\bibitem[1993]{gomez93} Gomez, M., Hartmann, L., Kenyon, S.~J., \& Hewett, R. 1993, \aj, 105, 1927

\bibitem[1969]{gr69} Gower, J.~C., \& Ross, G.~J.~S. 1969, Appl.\ Stat., 18, 54

\bibitem[2005]{haikala05} Haikala, L.~K., Harju, J., Mattila, K., \& Toriseva, M.
2005, \aap, 431, 149

\bibitem[2002]{hartmann02} Hartmann, L. 2002, \apj, 578, 914

\bibitem[1983]{hj83} Hoffman, R., \& Jain, A.~K. 1983, Pattern Recognition Letters, 1, 175

\bibitem[1996]{hb96} Hurt, R.~L., \& Barsony, M. 1996, \apj, 460, L45

\bibitem[2005]{jappsen05} Jappsen, A.-K., Klessen, R.~S., Larson, R.~B., Li, Y., \& Mac Low, M.-M.
       2005, \aap, 435, 611

\bibitem[2004]{kaas04} Kaas, A.~A., Olofsson, G., Bontemps, S., et al.\ 2004, \aap, 421, 623

\bibitem[1997]{klessen97} Klessen, R.~S. 1997, \mnras, 292, 11

\bibitem[2000]{klessen_burkert00} Klessen, R.~S., \& Burkert, A. 2000, \apjs, 128, 287

\bibitem[2001]{klessen+kroupa} Klessen, R.~S., \& Kroupa, P. 2001, \aap, 372, 105

\bibitem[1956]{kruskal56} Kruskal, J.~B., Jr. 1956, Proc.\ Amer.\ Math.\ Soc., 7, 48

\bibitem[2003]{lada_lada} Lada, C.~J., \& Lada, E.~A. 2003, \araa, 41, 57

\bibitem[1995]{larson95} Larson, R.~B. 1995, \mnras, 272, 213

\bibitem[2003]{larson03} Larson, R.~B. 2003, Rep.\ Prog.\ Phys., 66, 1651

\bibitem[1999]{lee_myers99} Lee, C.~W., \& Myers, P.~C. 1999, \apjs, 123, 233

\bibitem[2003]{luhman03} Luhman, K.~L., Stauffer, J.~R., Muench, A.~A., et al.
2003, \apj, 593, 109

\bibitem[2005]{li05} Li, Y., Mac Low, M.-M., \& Klessen, R.~S. 2005, \apj, 626, 823

\bibitem[1993]{marcelpoil93} Marcelpoil, R. 1993, Analytical Cellular Pathology, 5, 177

\bibitem[1992]{monaghan92} Monaghan, J.~J. 1992, \araa, 30, 543

\bibitem[1957]{prim57} Prim, R.~C. 1957, Bell Systems Tech.\ J., 36, 1389

\bibitem[1977]{rr77} Ripley, B.~D., \& Rasson, J.-P. 1977, J.\ Appl.\ Prob., 14, 483

\bibitem[2000]{robins00} Robins, V., Meiss, J.~D., \& Bradley, E. 2000, Physica D, 139, 276

\bibitem[SK04]{sk04} Schmeja, S., \& Klessen, R.~S. 2004, \aap, 419, 405 (SK04)

\bibitem[SKF05]{skf05} Schmeja, S., Klessen, R.~S., \& Froebrich, D. 2005, \aap, 437, 911 (SKF05)

\bibitem[2005]{stanke04} Stanke, T., Smith, M.~D., Gredel, R., \& Khanzadyan, T. 2005, \aap, in press
       (astro-ph/0511093)

\bibitem[2005]{vksb05} V{\'a}zquez-Semadeni, E., Kim, J., Shadmehri, M., \& Ballesteros-Paredes, J.\
2005, \apj, 618, 344


\end{thebibliography}
\end{document}